\documentclass[pra,showpacs,superscriptaddress,onecolumn]{revtex4}
\usepackage{amsfonts}
\usepackage{graphicx,amsmath,amssymb,epsfig}

\DeclareMathOperator{\sgn}{sgn}

\begin{document}

\title{Experimental analysis of decoherence in a CV bi--partite systems}
\author{D. Buono}
\email{danielabuono@yahoo.com}
\affiliation{Facolt\'{a} di Scienze MM. FF. NN., Universit\`{a} di Salerno, via Ponte don
Melillo - 84084 - Fisciano (SA)}
\author{G. Nocerino}
\email{gaetanocerino@libero.it}
\affiliation{Facolt\'{a} di Scienze MM. FF. NN., Universit\`{a} di Salerno, via Ponte don
Melillo - 84084 - Fisciano (SA)}
\author{A. Porzio}
\email{alberto.porzio@spin.cnr.it}
\affiliation{CNR -- SPIN, Napoli}
\affiliation{Dipartimento di Scienze Fisiche, Universit\`{a} \textquotedblleft Federico
II\textquotedblright , Complesso Univ. Monte Sant'Angelo, I-80126 Napoli,
Italy.}
\author{S. Solimeno}
\email{solimeno@na.infn.it}
\affiliation{Dipartimento di Scienze Fisiche, Universit\`{a} \textquotedblleft Federico
II\textquotedblright , Complesso Univ. Monte Sant'Angelo, I-80126 Napoli,
Italy.}

\begin{abstract}
Quantum properties are soon subject to decoherence once the quantum system
interacts with the classical environment. In this paper we experimentally
test how propagation losses, in a Gaussian channel, affect the bi--partite
Gaussian entangled state generated by a sub--threshold type--II opticla
parametric oscillator (OPO). Experimental results are discussed in terms of
different quantum markers, as teleportation fidelity, quantum discord and
mutual information, and continuous variable (CV) entanglement criteria. To
analyse state properties we have retrieved the composite system covariance
matrix by a single homodyne detector. We experimentally found that, even in
presence of a strong decoherence, the generated state never disentangles and
keeps breaking the quantum limit for the discord. This result proves that
the class of CV entangled states discussed in this paper would allow, in
principle, to realize quantum teleportation over an infinitely long Gaussian
channel.
\end{abstract}

\maketitle

\section{Introduction}

\textquotedblleft Quantumness\textquotedblright\ is very fragile and its
surviving over long distances and times is subtly interconnected to the
amount of noise coupling into the quantum system through its unavoidable
interaction with the classical environment. Among the different quantum
signatures, entanglement, a concept firstly introduced by Schr\"{o}dinger 
\cite{Schrodinger35} in response to the famous Einstein, Podolsky and Rosen
paper in 1935 \cite{EPR35}, probably represents the most fascinating one.
The conceptual consequences of entanglement are somehow subtle and deeply
affect many aspects of modern physics as the interaction between reality and
observer \cite{Wheeler84}, the actual limit for defining metrology standards 
\cite{Giovannetti06}, the limit to communication security \cite{Scarani09}.

The optical realization of such an intriguing quantum feature has gained, in
the last decades, a central role for realising complex quantum communication
protocols \cite{Braunstein05}. Among the different approaches the use of
bright beams, \textit{i.e.} optical modes carrying many photons, appears as
the most robust carrier of entangled information \cite{Ralph09} no matter if
the latter will be coded into single--photon--like states (discrete quantum
coding) or over the quadratures of the carrier beams (continuous valued
quantum coding) \cite{OBrien09}.

By interacting with the external world, a pure quantum state decoheres into
a mixture \cite{Serafini04} and its quantumness, if not completely death 
\cite{Yu09}, need to be restored or re--enhanced by suitable protocols \cite%
{Shor95}. As a matter of fact, decoherence limits both the attainable
protocol fidelity and the type of accessible protocols.

The aim of this paper is to discuss and experimentally analyse the effects
of the transmission over a lossy channel on the quantumness of bi--partite
Gaussian continuous variable (CV) optical entangled states. While these
effects on CV entangled states obtained by above threshold OPOs have been
already investigated \cite{Barbosa10}, we focus our analysis on the states
generated by a type-II sub--threshold OPO \cite{PRL09}.

The birth of an entangled state happens whenever a genuine quantum
correlation, \textit{i.e.} without classical analogue, is set between two
distinguishable quantum systems. This implies that the system wavefunction
cannot factorize into the product of wavefunctions of the single
sub--systems. While this definition is almost unique entanglement marks
itself into different properties of the quantum state of the whole system
and can be seen under different perspectives \cite{Wiseman07}. For this
reason, since the first experimental demonstration of entanglement \cite%
{Aspect81} several criteria and quantities have been introduced for
analyzing \textit{entanglementness} \cite{Reid89,Cavalcanti09}. Furthermore,
the definition of an entangled state is conceptually connected to pure
states \cite{Cavalcanti05}. The properties of these states are described by
wave functions and \textit{gedanken} experiments can be though for testing
different theoretical aspects but we all know that the states we can prepare
and manipulate in the laboratory are not pure but mixed. So that,
experimental entanglement tests, especially in CV regime, are described in
terms of density matrices rather than wavefunctions. Far from the ideal
concept of pure states and systems that do not interact with the environment
and/or observers, any quantum state undergoes to decoherence phenomena. In
optics the most common process leading to decoherence is the phase
insensitive loss of photons through diffusion and absorption mechanisms.
This process is described by a Lindblad equation \cite{Breur02} for the
evolution of the field operators that translates into a Master equation for
the state density matrix. Eventually, if the state is Gaussian, \textit{i.e.}
its Wigner function is Gaussian, it admits a full description in term of its
covariance matrix \cite{Simon87}.

The investigated CV entangled state represents one of the possible quantum
resources in CV teleportation protocols \cite{Vaidman94}. The experimental
data we present extend the analysis of Ref. \cite{Bowen03} discussing the
behaviour of the CV\ entangled system to the strong decoherence regime (up
99\% of loss). Moreover, we discuss in details the relationship between the
three different entanglement criteria used in the CV\ field linking them to
the teleportation fidelity and quantum discord, two possible quantum
signatures for evaluating the possibility of using this class of states in
quantum communication protocols \cite{Dakic12,Gu12}. The discussion on the
experimental results is preceded by a detailed theoretical foreword aimed at
giving a physical insight the different criteria and quantities actually
used for classifying CV\ entanglement. With this paper we prove that the
state we are able to generate never disentangles. This rather
counter--intuitive result has also been shown theoretically for the case of
qubits states, showing a purity at the generation stage above a threshold
value \cite{CavalcantiOPEX09}. Moreover, the state we analyse can be used as
the quantum resource for an effective quantum teleportation, of a coherent
state, for any value of the transmission loss although the teleportation
fidelity slowly reduce to its quantum limit.

Optical entangled states can be obtained in non--linear processes such as
parametric amplifiers that, depending on their operating regime, can prepare
either single photon \cite{Kwiat95} or continuous variable Gaussian
entangled states \cite{Ou92,Bowen02,Laurat05,Villar05}. In the latter case
the non--linear medium is allocated in a optical cavity and the OPOs, below
the oscillation threshold and in a semiclassical approach, are described by
bilinear Hamiltonians so realising the paradigm for Gaussian state
generation \cite{Collett84}. In particular below threshold a single
continuous wave OPO, generating squeezing in a fully degenerate operation 
\cite{Wu86}, can give raise to a pair of bright CV entangled beams in the
non--degenerate case \cite{Laurat05,Villar05,Su06,Jing06,Keller08}. Both the
cases lead to states that represent robust resources for implementing
different quantum communication tasks \cite{Weedbrook12}.

In this paper we experimental discuss the behaviour, under strong loss, of a
bright bi--partite CV Gaussian entangled state outing a sub--threshold
type--II OPO. In particular, we analyse, for this state, the behaviour of
different quantumness and entanglement markers in order to discuss the limit
at which this state can be transmitted before loosing its quantum ability of
being employed in quantum communication protocols. More in details, we
discuss: state purity, mutual information, quantum discord and three
different entanglement criteria: Peres--Horodechi--Simon \cite%
{Peres96,Simon00}, Duan \cite{Duan00}, and EPR--Reid \cite{Reid89}. In
addition we relate these criteria to the teleportation fidelity $\mathcal{F}$%
, \textit{i.e.} the state ability of overcoming the quantum limit in a
teleportation protocol for a coherent state.

These states, already successfully generated in the laboratories \cite%
{Ou92,Laurat05,PRL09}, are Gaussian, \textit{i.e.} they are completely
characterized by their covariance matrix, namely, by the first and second
moments of their quadratures so that they can be easily characterized by a
single homodyne detector \cite{JOB05}.

The paper is organized as follows: first, in Sect. \ref{Sect:state} a
general overview over the state generated by a sub--threshold OPO is given
and the model for its evolution in a lossy Gaussian channel is discussed. In
Sect. \ref{Sect:markers} the different entanglement markers are introduced
in terms of the state covariance matrix. The section includes a detailed
discussion on the relationship between the different entanglement criteria
and quantum markers for CV Gaussian states. Next, in Sect. \ref{Sect:evolved}
the quantum markers' evolution in term of channel transmission is given. A
brief description of the experimental apparatus, given in Sect. \ref%
{Sect:Experiment}, precedes the discussion, in Sect. \ref{Sect:Results}, of
our experimental findings. Eventually, in Sect. \ref{Sect:conclusions},
conclusions are drawn.

\section{Gaussian CV bi--partite system\label{Sect:state}}

A sub--threshold frequency degenerate type--II OPO generates a bi--partite
state made of the two downconverted cross--polarized beams \cite{Drummond90}%
. These states are Gaussian \textit{i.e. }the Wigner function ($W$) of the
state $\rho $, expressed in terms of field quadratures is Gaussian. In our
case, we can assume that the average values of quadratures to be zero \cite%
{PRL09} so that we can write $W$ as%
\begin{equation}
W\left( \mathbf{K}\right) =\frac{\exp \left[ -\frac{1}{2}\mathbf{K}^{\top }%
\mathbf{\sigma }^{-1}\mathbf{K}\right] }{\pi ^{2}\sqrt{\text{Det}\left[ 
\mathbf{\sigma }\right] }}~,  \label{Wigner0}
\end{equation}%
where $\mathbf{K\equiv }\left( X_{1},Y_{1},X_{2},Y_{2}\right) ^{\top }$ is
the vector of the field quadratures (hereafter, whenever not necessary, we
will omit the "$~\widehat{}~$" on operators so that: $X=\left( a^{\dag
}+a\right) /\sqrt{2}$ and $Y=i\left( a^{\dag }-a\right) /\sqrt{2}$) for mode 
$1$ and $2$ respectively.

A Gaussian state can be fully characterized by the matrix of the second
order statistical moments of the field quadratures, \textit{i.e.} the
covariance matrix (\textbf{CM}) $\mathbf{\sigma }$, whose elements are
defined by $\sigma _{kh}\equiv \frac{1}{2}\left\langle \left\{
K_{k},K_{h}\right\} \right\rangle -\left\langle K_{k}\right\rangle
\left\langle K_{h}\right\rangle $ where $\left\{ K_{k},K_{h}\right\}
=K_{k}K_{h}+K_{h}K_{k}$. Moreover, for a bi--partite field state, $\mathbf{%
\sigma }$ can be always given in the standard form by local symplectic
operations%
\begin{equation}
\mathbf{\sigma =}\left( 
\begin{array}{cc}
\mathbf{\alpha } & \mathbf{\gamma } \\ 
\mathbf{\gamma }^{\top } & \mathbf{\beta }%
\end{array}%
\right) =\left( 
\begin{array}{cccc}
n & 0 & c_{1} & 0 \\ 
0 & n & 0 & c_{2} \\ 
c_{1} & 0 & m & 0 \\ 
0 & c_{2} & 0 & m%
\end{array}%
\right) ~,  \label{CM initial}
\end{equation}%
where $\mathbf{\alpha },$ $\mathbf{\beta }$ and $\mathbf{\gamma }$ are $%
2\times 2$ real matrices, representing respectively the self and mutual
correlations matrices. The quantities, $n$, $m$, $c_{1}$ and $c_{2}$ are
determined by the four local symplectic invariants $I_{1}\equiv \det (%
\mathbf{\alpha })=n^{2}$, $I_{2}\equiv \det (\mathbf{\beta })=m^{2}$, $%
I_{3}\equiv \det (\mathbf{\gamma })=c_{1}c_{2}$, $I_{4}\equiv \det ({%
\boldsymbol{\sigma }})=\left( nm-c_{1}^{2}\right) \left( nm-c_{2}^{2}\right) 
$. We note that the experimental matrices we have measured and that will be
discussed in Sect. \ref{Sect:Results} are all in the above standard form. As
a matter of fact, a sub--threshold type--II OPO, due to the symmetry of its
Hamiltonian, can only produce states of this kind. So that, hereafter,
whenever we refer to \textbf{CM} we will intend a covariance matrix in the
standard form of Eq. (\ref{CM initial}). Moreover, at the time of their
birth, states produced in an OPO show $n=m$ and $c_{1}=-c_{2}$, and the
matrix is called symmetric and represents a bi--partite state where the
energy is equally distributed between the two modes.

${\boldsymbol{\sigma }}$ is a \textit{bona fide} \textbf{CM} \textit{i.e.}
it describes a physical state iff 
\begin{equation}
{\boldsymbol{\sigma }}+\frac{i}{2}\boldsymbol{\omega }\oplus \boldsymbol{%
\omega \geq }0~,  \label{Heisenberg principle}
\end{equation}%
where $\boldsymbol{\omega \equiv }\left( 
\begin{array}{cc}
0 & 1 \\ 
-1 & 0%
\end{array}%
\right) $. The above condition can be written in terms of the four
symplectic invariants%
\begin{equation}
I_{1}+I_{2}+2I_{3}\leq 4I_{4}+\frac{1}{4}~.
\label{Heis rel. sympl invariant}
\end{equation}%
The above inequalities are equivalent to the Heisenberg uncertainty
principle for a two--mode state and to ask covariance matrix positivity.

The \textbf{CM} is also characterized by its symplectic eigenvalues%
\begin{equation}
d_{\pm }=\sqrt{\frac{I_{1}+I_{2}+2I_{3}\pm \sqrt{\left(
I_{1}+I_{2}+2I_{3}\right) ^{2}-4I_{4}}}{2}}~.  \label{symplectic}
\end{equation}%
Inequality (\ref{Heis rel. sympl invariant}) assumes a simple form in term
of $d_{-}$:%
\begin{equation*}
d_{-}\geq \frac{1}{2}~.
\end{equation*}

We also note that a Gaussian pure state is a minimum uncertainty state and
that the \textbf{CM} relative to a pure state necessary has $\det ({%
\boldsymbol{\sigma }})=I_{4}=1/16$ (see Eq. (\ref{purity}) below) so that
the case $c_{1}=-c_{2}=c$ implies $n=m$, to ensure a \textit{bona fide} 
\textbf{CM,} and for a pure fully symmetric state%
\begin{equation}
c=\sqrt{n^{2}-1/4}~.  \label{c_pure}
\end{equation}%
Moreover, for mixed fully symmetric states, $c\leq \sqrt{n^{2}-1/4}$ with
the inequality saturated only by pure states. Hereafter we will indicate
these fully symmetric states as \textit{diagonal}. The meaning of this name
will be clarified in Sect. \ref{Sect:markers}.

\subsection{Evolution in a lossy Gaussian channel\label{Sect:evolution}}

Decoherence indicates the detrimental effect on a quantum system that
stochastically interacts with the external world. The state of the system,
initially pure, becomes mixed. In the density matrix language it translates
into the fact that while a pure state is represented by an idempotent
density matrix $\rho $ ($\rho ^{2}=\rho $) this is not true for a decohered
mixed state. The lossy transmission of an arbitrary optical quantum state
between two sites is an irreversible decoherence process that can be
described by using the open systems approach \cite{Breur02}. In this
approach the environment is seen as a reservoir (thermal bath) made up of
infinite modes at thermal equilibrium. The Kossakowski-Lindblad equation
describes the time evolution in a noisy quantum channel of a multi--mode
quantum state $\rho _{S}$. This model is valid if the system $S$ and the
reservoir $R$ satisfy the following general conditions:

\begin{itemize}
\item \textit{weak coupling (Born approximation)}\textbf{\ --} the coupling
between system and environment is so that the density matrix $\rho _{R}$ of
the environment is negligibly influenced by the interaction (the thermal
bath state is stationary). This approximation allows to write the state $%
\rho _{SR}\left( t\right) $ of the global system as $\rho _{SR}\left(
t\right) \approx $ $\rho _{S}\left( t\right) \otimes \rho _{R}$;

\item \textit{Markovianity -- }there are not memory effects neither on the
system and the reservoir. This approximation implies that the time scale $%
\tau $ over which $\rho _{S}\left( t\right) $ changes appreciably under the
influence of the bath is large compared to the time scale $\tau _{R}$ over
which the bath forgets about its past, $\tau \gg \tau _{R}$.

\item \textit{Secularity (rotating wave approximation) -- }the typical time
scale $\tau _{S}$ of the intrinsic evolution of the system $S$ is small
compared to the relaxation time $\tau $. For an optical field it implies
that the reservoir reacts to the average field and not to its instantaneous
value.
\end{itemize}

The evolution of an arbitrary two--mode Gaussian state in a noisy channel,
at thermal equilibrium, can be translated into the formalism of the Wigner
function obtaining the following Fokker-Planck equation for $W\left( \mathbf{%
K}\right) $ (see Eq.(\ref{Wigner0})) \cite{Serafini04},%
\begin{equation}
\partial _{t}W\left( \mathbf{K},t\right) =\frac{\Gamma }{2}\left( \partial _{%
\mathbf{K}}^{\top }\mathbf{K+}\frac{1}{2}\nabla _{\mathbf{K}}^{2}\right)
W\left( \mathbf{K}\right) ~,  \label{Fokker-Planck}
\end{equation}%
where $\Gamma $ is the damping rate of the channel for both modes, while $%
\partial _{\mathbf{K}}^{\top }=\left( \partial _{X_{1}},\partial
_{Y_{1}},\partial _{X_{2}},\partial _{Y_{2}}\right) $ and $\nabla _{\mathbf{K%
}}^{2}=\partial _{\mathbf{K}}^{\top }\partial _{\mathbf{K}}=\partial
_{X_{1}}^{2}+\partial _{Y_{1}}^{2}+\partial _{X_{2}}^{2}+\partial
_{Y_{2}}^{2}$.

The evolution of Eq.(\ref{Fokker-Planck}) preserves the Gaussian character
of the initial state and in terms of the \textbf{CM} $\mathbf{\sigma }$ reads%
\begin{equation}
\mathbf{\sigma }(t)\mathbf{=}\left( 1-e^{-\Gamma t}\right) \frac{1}{2}%
\mathbb{I}+e^{-\Gamma t}\mathbf{\sigma }(0),  \label{CM evolution}
\end{equation}%
where $\mathbf{\sigma }(0)$ is the covariance matrix at $t=0$ and $\mathbb{I}
$ is the $4\times 4$ identity matrix ($\frac{1}{2}\mathbb{I}$ is the vacuum
state covariance matrix setting the standard quantum limit SQL).

This form is in all equal to the effects of a fictitious beam--splitter (BS)
that mimics the channel losses and couples into the system the vacuum
quantum noise through its unfilled port. Being $U_{k}\left( \zeta \right)
=\exp \left\{ \zeta \left( a_{k}^{\dag }v_{k}-v_{k}^{\dag }a_{k}\right)
\right\} $ the SU(2) transformation induced by the BS on the $k$--mode ($%
k=1,~2$, with $v_{k}$ the modal operator for the vacuum) and $T=e^{-\Gamma
t} $ the power transmission of the beam splitter ($\tan \zeta =\sqrt{\left(
1-T\right) /T}$), the above equation becomes:%
\begin{equation}
\mathbf{\sigma }_{T}\mathbf{=}\left( 1-T\right) \frac{1}{2}\mathbb{I}+T%
\mathbf{\sigma }_{1}~.  \label{CM post BS}
\end{equation}%
In this form we can drop the temporal dependence and label the \textbf{CM}
of the initial state as $\mathbf{\sigma }_{T=1}\equiv \mathbf{\sigma }_{1}$.
For states in the form of Eq. (\ref{CM initial}) the different elements
evolve as%
\begin{eqnarray}
n_{T} &=&\frac{1}{2}+T\left( n-\frac{1}{2}\right)  \notag \\
m_{T} &=&\frac{1}{2}+T\left( m-\frac{1}{2}\right)  \notag \\
c_{1(2),T} &=&c_{1(2)}T  \label{CM_elements_evolution}
\end{eqnarray}

For an infinite transmission channel $T\rightarrow 0$ implies $\mathbf{%
\sigma }_{T}\rightarrow \frac{1}{2}\mathbb{I}$ \textit{i.e.} the fully
decohered \textbf{CM} represents a two--mode vacuum state.

\section{Quantum markers and entanglement\label{Sect:markers}}

The quantumness of a bi--partite \textbf{CV} state can be tested by two
classes of markers. The first is intimately related to the state itself and
includes inequalities that fix bounds for distinguishing among entangled and
un--entangled states. The second has been translated into the quantum
context from the classical information theory and is related to the amount
of quantum information carried by the state. The latter includes
quantitative measures such as mutual information, von Neumann entropy,
quantum discord and teleportation fidelity. The former is made of
entanglement criteria usually named by the authors that have theoretically
found them. They are the PHS (Peres-Horodecki-Simon) \cite{Peres96,Simon00},
Duan\cite{Duan00}, EPR--Reid \cite{Reid89} criteria. Before going into
details it is useful to start by introducing the state purity as an extra
marker that tells how far the analysed state is from a pure one.

\subsubsection{State purity\label{Sect:purity}}

Every quantum information protocol and every measurement scheme must take
into account the fact that pure states are inevitably corrupted by the
interaction with the environment and, therefore, in real experiments only
mixed states are available. The purity $\mu =Tr[\rho ^{2}]$ ($\mu <1$ for
mixed state, $=1$ for pure) may be expressed as a function of the \textbf{CM}
(\ref{CM initial}) being%
\begin{equation}
\mu =\frac{1}{4\sqrt{\text{det}\left[ \mathbf{\sigma }\right] }}~.
\label{purity}
\end{equation}%
Intimately related to the purity, it is possible to introduce the von
Neumann entropy $S\left( \rho \right) $. It is zero for pure states while
strictly positive for mixed ones. In the case of the two--mode Gaussian
state the entropy $S\left( \rho \right) $ can be written in terms of its
covariance matrix symplectic eigenvalues $d_{\pm }$ (see Eq. (\ref%
{symplectic})) as $S\left( \rho \right) =f\left( d_{+}\right) +f\left(
d_{-}\right) $ \cite{Serafini04JPB} where 
\begin{equation}
f\left( x\right) =\left( x+1/2\right) \log \left( x+1/2\right) -\left(
x-1/2\right) \log \left( x-1/2\right) ~.  \label{Funct_f}
\end{equation}%
We note that both $\mu $ and $S\left( \rho \right) $ are invariant under
symplectic transformations.

\subsection{Entanglement criteria}

Although it has not been found a general solution to the problem of a
quantitative measure of entanglement for mixed states, there exist necessary
and sufficient conditions to asses whether a given state is entangled or
not.\ These criteria provide a test for entanglement, and can be employed
for studying the behaviour of entanglement in a quantum transmission channel.

\subsubsection{The Peres-Horodecki-Simon (PHS) criterion\label{Sect:PHS}}

A bi--partite quantum state is separable iff its density operator can be
written as a convex combination of the tensor product of density operators
relative to the two different sub--systems \cite{Werner89}%
\begin{equation}
\rho =\sum_{j}p_{j}\rho _{j1}\otimes \rho _{j2},  \label{SepSt}
\end{equation}%
where $\sum_{j}p_{j}=1$ while $\rho _{ji}$ $i=1,2$ are the density matrices
of subsystems $1$ and $2$. By performing a partial transposition (\textit{%
i.e.} transposition of the density matrix with respect to only one of the
two Hilbert subspaces) $\rho $ transforms into $\rho _{PT}$. In the case $%
\rho $ is a separable state, $\rho _{PT}$ should still represent a physical
state so that $\rho _{PT}$ must be a non--negative density operator. On the
contrary if $\rho _{PT}$ no more represents a physical state the system does
not admit the form (\ref{SepSt}). Thus, all separable states have a
non-negative partially transposed density operator. From this consideration
it is possible to deduce a necessary condition for separability or viceversa
a sufficient condition for entanglement. In view of this the criterion is
sometime referred to as the \textit{ppt} criterion (\textit{positivity}
under \textit{partial transposition}) \cite{Peres96,Simon00}. In the
following we prefer to indicate it as the PHS criterion. It can be proven
that it becomes a necessary and sufficient condition for Gaussian states 
\cite{Simon00}. From the \textbf{CM} point of view, partial transposition
implies a sign flip for $I_{3}$ so that PHS criterion has a simple
expression in terms of \textbf{CM} elements: a bi--partite Gaussian state is
separable iff 
\begin{equation}
n^{2}+m^{2}+2\left\vert c_{1}c_{2}\right\vert -4\left( nm-c_{1}^{2}\right)
\left( nm-c_{2}^{2}\right) \leq \frac{1}{4}\text{ },  \label{PHScr}
\end{equation}%
and it is entangled otherwise.\textbf{\ }We also note that the PHS criterion
is invariant under symplectic transformations.

PHS criterion relies on the possibility of describing independently the two
subsystems. In phase space, transposition is defined as \textquotedblleft
time reversal\textquotedblright\ (or mirror reflection), which is given by a
change of sign of out--of--phase quadrature (usually indicated as $Y$).
Partial transposition is, therefore, a \textquotedblleft local time
reversal\textquotedblright\ which inverts the $Y$ quadrature of only one
subsystem. If any \textit{true quantum} correlation is set between $Y_{1}$
and $Y_{2}$ a sign flip on $Y_{1}$ (or $Y_{2}$) will affect the sign of $%
c_{1}$ (or $c_{2}$) in Eq. (\ref{CM initial}) making $\rho _{PT}$ no more
physical (see Eq. (\ref{Heis rel. sympl invariant})).

\subsubsection{The Duan criterion}

For every bi--partite state there exists a pair of EPR--like conjugate
operators defined by%
\begin{equation}
\hat{u}=\left\vert a\right\vert \hat{x}_{1}+\frac{1}{a}\widehat{x}_{2}\text{
and }\widehat{v}=\left\vert a\right\vert \widehat{p}_{1}-\frac{1}{a}\widehat{%
p}_{2}~,  \label{EPRop}
\end{equation}%
with $a$ an arbitrary non--zero real number and $\left[ \hat{x}_{j},\hat{p}%
_{j^{\prime }}\right] =\frac{i}{2}\delta _{jj^{\prime }}$ ($j,j\prime =1,~2$%
) and where subscript refers to the $1$ ($2$) subsystem. Then, the birth of
non--classical correlation between sub--systems $1$ and $2$ will lead to
states for which the variance of EPR--like operators will reduce below the
standard quantum limit (SQL). By calculating the total variance of such a
pair of operators on $\rho $, a separable state of the form of Eq. (\ref%
{SepSt}), it can be proven \cite{Duan00} that 
\begin{equation}
\left\langle (\Delta \widehat{u})^{2}\right\rangle _{\rho }+\left\langle
(\Delta \widehat{v})^{2}\right\rangle _{\rho }\geq a^{2}+\frac{1}{a^{2}}~,
\label{DUAN}
\end{equation}%
setting a lower bound for separable states. Contrarily to the PHS criterion
(see Sect. \ref{Sect:PHS}) inequality (\ref{DUAN}) has been formulated only
as a necessary condition for separability so that it is a sufficient
condition for entanglement of a generic CV\ Gaussian state.

As shown in \cite{Duan00} it becomes a necessary and sufficient condition
for entangled CV\ Gaussian states. The sufficient and necessary condition
can be expressed in terms of the covariance matrix elements iff the matrix
itself is expressed in the form of Eq. (10) of Ref. \cite{Duan00}%
\begin{equation}
\mathbf{\sigma }=\left( 
\begin{array}{cccc}
n_{1} & 0 & c_{1} & 0 \\ 
0 & n_{2} & 0 & c_{2} \\ 
c_{1} & 0 & m_{1} & 0 \\ 
0 & c_{2} & 0 & m_{2}%
\end{array}%
\right) ~,  \label{CM_Duan}
\end{equation}%
with the matrix elements satisfying the constrains (11a) and (11b) of Ref. 
\cite{Duan00} that, for the SQL equal to $\frac{1}{2}$, read%
\begin{eqnarray}
\frac{n_{1}-1/2}{m_{1}-1/2} &=&\frac{n_{2}-1/2}{m_{2}-1/2}~,  \notag \\
\left\vert c_{1}\right\vert -\left\vert c_{2}\right\vert &=&\sqrt{\left(
n_{1}-1/2\right) \left( m_{1}-1/2\right) }-\sqrt{\left( n_{2}-1/2\right)
\left( m_{2}-1/2\right) }~.  \label{Duan_conditions}
\end{eqnarray}%
In this case the EPR operators pair of Eq. (\ref{EPRop}) are rewritten as%
\begin{equation*}
\widehat{u}=a_{0}\widehat{x}_{1}+%
\sgn(c_{1})%
\frac{1}{a_{0}}\widehat{x}_{2}\text{ and }\widehat{v}=a_{0}\widehat{p}_{1}-%
\sgn(c_{2})%
\frac{1}{a_{0}}\widehat{p}_{2}~,
\end{equation*}%
where $a_{0}=\sqrt{\frac{m_{1}-1/2}{n_{1}-1/2}}=\sqrt{\frac{m_{2}-1/2}{%
n_{2}-1/2}}$ with the sufficient and necessary Duan criterion given by%
\begin{equation}
a_{0}^{2}\left( n_{1}+n_{2}-1\right) +\frac{m_{1}+m_{2}-1}{a_{0}^{2}}%
-2\left( \left\vert c_{1}\right\vert -\left\vert c_{2}\right\vert \right)
<0~.  \label{Duan_necessary}
\end{equation}%
We note that, as proved in Ref. \cite{Duan00}, any Gaussian state can be
transformed into the form (\ref{CM_Duan}) by local linear unitary Bogoliubov
operations, \textit{i.e.} by acting independently on one or both the
subsystems by applying local squeezing and/or rotations.

In a more general fashion it is possible to write the sufficient but not
necessary condition of Eq. (\ref{DUAN}) for a \textbf{CM} in the standard
form of Eq. (\ref{CM initial}) as%
\begin{equation*}
\left( 2n-1\right) a^{2}+\frac{\left( 2m-1\right) }{a^{2}}-2\left(
c_{1}-c_{2}\right) <0~,
\end{equation*}%
where $a$ can be set to $a^{2}=\sqrt{\frac{m-1/2}{n-1/2}}$ to minimize the
left hand side of the inequality:%
\begin{equation}
\sqrt{\left( 2n-1\right) \left( 2m-1\right) }-\left( c_{1}-c_{2}\right) <0~.
\label{DUANcr}
\end{equation}%
We note that, while for symmetric states ($m=n$) $\left\vert a\right\vert =1$
and the EPR pair consists of two orthogonal field quadratures, this is not
true, in general.

The Duan criterion is strictly related to the Heisenberg principle for the
single sub--system. If the state is separable the indeterminacy on a single
operator is disjoint from the indeterminacy of the twin operator on the
second sub--system; so that the total indeterminacy cannot violate the
Heisenberg limit. This has nothing to do with conditional measurement and
with the possibility of gaining information on one sub--system measuring the
other. As we will see this approach leads to a stricter criterion: the
so--called EPR "\textit{Reid}" criterion.

\subsubsection{The EPR "Reid" criterion}

A stronger bound can be found by analysing a bi--partite state under the
shadow of conditional measurements. This concept descends directly from the
original EPR \textit{gedanken }experiment \cite{EPR35}. For this reason it
is usually indicated as the EPR criterion and was firstly introduced by Reid
in 1989 \cite{Reid89}, in the very early days of quantum information. It
describes the ability to deduce the expectation value of an observable on a
sub--system by measuring the EPR companion observable on the second
sub--system. The EPR "Reid" criterion sets, as the Duan one in its generic
form (\ref{DUANcr}), only a sufficient condition for assessing entanglement.

Mathematically this criterion can be deduced by calculating the conditional
variance for an observable on sub--system $1$ given the result of a
measurement on sub--system $2$ and comparing it with the standard quantum
limit.

For Gaussian states it can be written in terms of \textbf{CM} elements: 
\begin{equation}
n^{2}\left( 1-\frac{c_{1}^{2}}{nm}\right) \left( 1-\frac{c_{2}^{2}}{nm}%
\right) <\frac{1}{4}~.  \label{EPRcr}
\end{equation}%
We note that, being based on conditional variances (and thus on conditional
states) this last criterion is not symmetric under the exchange of the two
sub-systems. This asymmetry reveals itself in a asymmetric sensitivity to
loss \cite{Reid09}. Then, the criterion itself can be recast if sub--system $%
1$ is measured and the conditional variance on $2$ is given%
\begin{equation*}
m^{2}\left( 1-\frac{c_{1}^{2}}{nm}\right) \left( 1-\frac{c_{2}^{2}}{nm}%
\right) <\frac{1}{4}~.
\end{equation*}%
The two definitions of the EPR\ criterion can make it ambiguous if one of
the relations are not satisfied. This is not the case of balanced systems ($%
m=n$). Moreover, it can be proved that no pure state can asymmetrically
violate the EPR criterion. At the same time this asymmetry can be useful for
revealing eavesdropping in one--sided device--independent QKD schemes \cite%
{Branciard12}.

It is easy to see that the above two expressions for the EPR criterion are
invariant for symplectic transformations like the PHS one (see Eq. (\ref%
{PHScr})).

The EPR criterion is necessary and sufficient for assessing the
\textquotedblleft steering\textquotedblright\ form of nonlocality as
recently discussed in Refs. \cite{Wiseman07,Cavalcanti09}. This concept is
very recently gaining much attention as a peculiar consequence of
entaglement \cite{Midgley10}. It is interesting to note that the concept of
conditional variance, used for quantum steering, was used for assessing
state preparation ability in quantum demolition experiments \cite{Holland90}%
. In that case steering happens between the probe and the signal systems.

\subsubsection{Witnesses}

All the above criteria (\ref{PHScr}), (\ref{DUANcr}), and (\ref{EPRcr})
cannot be used other than as bounds. They are not suitable for measuring
entanglement in a quantitative way. Quantifying entanglement for mixed
states is a complicated issue still under discussion. There isn't, to date,
a single, universal measure that quantifies the entanglement for a mixed
state.

In discussing the experimental results in Sect. \ref{Sect:Results}, we will
make use of three different witnesses, one each for the above criteria, in
order to evaluate their behaviours under lossy transmission:%
\begin{eqnarray}
w_{PHS} &=&4\left( nm-c_{1}^{2}\right) \left( nm-c_{2}^{2}\right) +\frac{1}{4%
}-\left( n^{2}+m^{2}\right) -2\left\vert c_{1}c_{2}\right\vert ~,  \notag \\
w_{DUAN} &=&2\sqrt{\left( n-\frac{1}{2}\right) \left( m-\frac{1}{2}\right) }%
-\left( c_{1}-c_{2}\right) ~,  \notag \\
w_{EPR} &=&n^{2}\left( 1-\frac{c_{1}^{2}}{nm}\right) \left( 1-\frac{c_{2}^{2}%
}{nm}\right) -\frac{1}{4}~.  \label{witnesses}
\end{eqnarray}

The three $w$'s don't satisfy the requirements for being a measure of
entanglement. For example they don't verify the basic axiom stating that a
good measure should be equal to $0$ for any separable state \cite{Vedral98}.

In summary, for a Gaussian bi--partite state the three criteria (see Eqs. (%
\ref{PHScr}), (\ref{DUANcr}), and (\ref{EPRcr})) reduce to%
\begin{equation}
\rho \text{ is entangled }\left\{ 
\begin{array}{l}
\iff w_{PHS}<0 \\ 
\impliedby \left\{ 
\begin{array}{c}
w_{DUAN}<0 \\ 
w_{EPR}<0%
\end{array}%
\right.%
\end{array}%
\right. ~.  \label{Ent_cr}
\end{equation}

However, once the state $\rho $ is entangled $w_{PHS}$, $w_{DUAN}$ and $%
w_{EPR}$ provide suitable markers for evaluating how far the state is from
being separable, somehow measuring the robustness of the entanglement.

Eventually we note that for \textit{diagonal }fully symmetric states ($n=m$
and $c_{1}=-c_{2}=c$ in Eq. (\ref{CM initial})) $w_{PHS}$, $w_{DUAN}$ and $%
w_{EPR}$ read%
\begin{eqnarray*}
w_{PHS} &=&4\left( n^{2}-c^{2}\right) ^{2}+\frac{1}{4}-2n^{2}-2c^{2}~, \\
w_{DUAN} &=&\left( n-\frac{1}{2}\right) -c~, \\
w_{EPR} &=&n^{2}\left( 1-\frac{c^{2}}{n^{2}}\right) ^{2}-\frac{1}{4}~;
\end{eqnarray*}%
and the two bounds ($c>n-1/2$) for $w_{PHS}$ and $w_{DUAN\text{ }}$coincide
while the bound for $w_{EPR}$ is $c>\sqrt{n\left( n-\frac{1}{2}\right) }$ so
that the EPR criterion is stricter than the PHS and Duan ones for any
allowed value of $n$.

\subsection{Quantifying entanglement}

The quality of an entangled state can be, indeed, evaluated,\ operatively,
by looking at the effectiveness such a state could, in principle, give in
quantum communication protocols. To this end we use $\mathcal{F}$, the
fidelity of CV teleportation of a coherent state, as a benchmark. In fact, $%
\mathcal{F}$ for a Gaussian resource depends only on the entanglement
quality of the resource itself. Using as resource a Gaussian two-mode state
described by the \textbf{CM} in Eq.\textbf{\ }(\ref{CM initial}), $\mathcal{F%
}$ is given by \cite{Pirandola06}%
\begin{equation}
\mathcal{F}=\left( 1+m+n-2c_{1}\right) ^{-\frac{1}{2}}\left(
1+m+n+2c_{2}\right) ^{-\frac{1}{2}}~.  \label{Fidelity_th}
\end{equation}%
It is possible to show that a teleportation protocol of a coherent state,
fully based on classical strategies, provides $\mathcal{F\leq }1/2$. Then, $%
\mathcal{F>}1/2$ implies that the resource is entangled. So that, the value
of $\mathcal{F}$ becomes an indicator of the quality of the entanglement. We
note that for \textit{diagonal} states ($n=m$ and $c_{1}=-c_{2}=c$) $%
\mathcal{F}=(1+2n-2c)^{-1}=\left( 2+w_{DUAN}\right) ^{-1}$. In this
particular case, perfect fidelity would be obtained for $w_{DUAN}=-2$ that
would imply an unphysical \textbf{CM} so that Gaussian resources, as the
state produced by OPOs, cannot guarantee perfect teleportation \cite%
{DellAnno07}. Moreover, $\mathcal{F}>1/2$ gives $c>n-1/2$ and coincides with
both the Duan and PHS bounds (see the end of the previous subsection) for
such \textit{diagonal }states. We note that, similarly to the Duan (Eq. (\ref%
{DUANcr})) but contrarily to the PHS and EPR criteria (Eqs. (\ref{PHScr})
and (\ref{EPRcr})) $\mathcal{F}$ is not invariant under symplectic
transformations.

\begin{figure}[tph]
\includegraphics[width=0.48\textwidth]{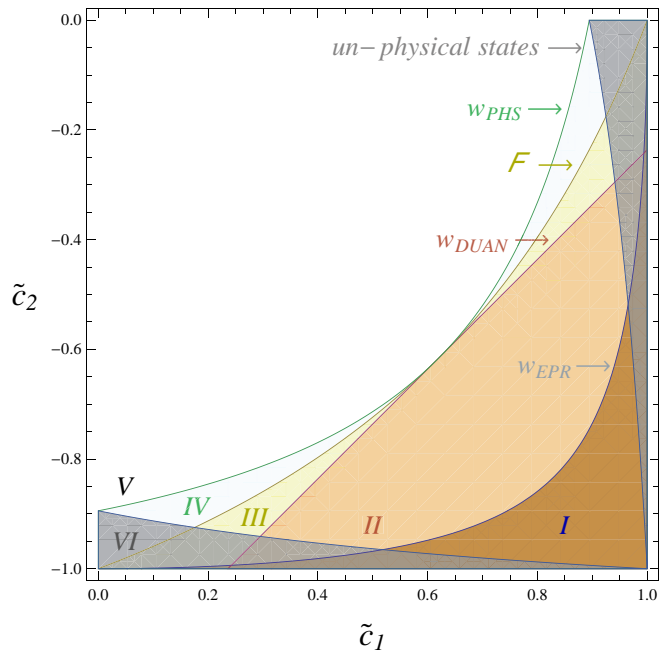}
\caption{Region plot (\textit{color online}) of the different entaglement
witnesses of Eqs. (\protect\ref{witnesses}) and teleportation fidelity (Eq. (%
\protect\ref{Fidelity_th})) as an entanglement marker. The \textit{light gray%
} (labelled with (VI)) areas indicate un-physical \textbf{CM}s (\textit{i.e.}
violating inequality (\protect\ref{Heis rel. sympl invariant})). The
different criteria show different regions of entanglement (see text for
details).}
\label{Fig:criteria}
\end{figure}

As already mentioned, only the PHS criterion can be written for a generic 
\textbf{CM} as a sufficient and necessary condition while the Duan and the
EPR ones set only sufficient bounds. In order to discuss the relations
between the different bounds and the teleportation fidelity, as an
entanglement marker, it is possible to draw a region plot, similar to the
one drawn in Ref. \cite{Barbosa11} for a different case. In Fig. \ref%
{Fig:criteria}, we have visualized the different bounds set by the three
entanglement witnesses of Eq. (\ref{witnesses})) and the region for which $%
\mathcal{F}>1/2$ (see Eq. (\ref{Fidelity_th})). The plot has been computed
considering a \textbf{CM} in the form of Eq.\textbf{\ }(\ref{CM initial})
with $m=n$ (balanced system). The axes report the value of $\tilde{c}_{1}$
(below) and $\tilde{c}_{2}$ (left), the two correlation terms of the
covariance matrix are normalized to $c_{MAX}=\sqrt{n^{2}-1/4}$ so that $%
\tilde{c}_{1}=-\tilde{c}_{2}=1$ represents a pure maximally entangled state (%
\textit{i.e.} the state showing the maximum quantum correlation for a given
total energy of the system). We note that fully symmetric states, that we
indicated as \textit{diagonal} states, lay on the plot diagonal (top--left
to bottom--right). These states, besides their symmetry, can be obtained at
the end of a lossy propagation, described by Eq. (\ref{CM evolution}), of an
initially pure state. On the contrary, states outside the diagonal, having $%
\tilde{c}_{1}\neq -\tilde{c}_{2}$, cannot be obtained by propagating pure
states. As a matter of fact, CV entangled states produced by type--II OPO,
show, in view of the symmetry in the interaction Hamiltonian, $\tilde{c}%
_{1}=-\tilde{c}_{2}$. This is not true for CV entangled states obtained by
mixing the outputs of two independent type--I OPOs. In such a case the two
fields have disjoint Hamiltonians and the symmetry is broken.

The \textit{light gray} (labelled with (\textit{VI})) areas indicate
un--physical states \textit{i.e.} \textbf{CM}s violating inequality (\ref%
{Heis rel. sympl invariant}).

\textit{Diagonal} states satisfy the conditions (see Eq. (\ref%
{Duan_conditions})) for which the Duan criterion becomes also necessary so
that the coincidence between the Duan and the PHS bounds, along the
diagonal, is not a surprise. Being both necessary and sufficient they
coincide. For these \textit{diagonal} states entanglement (seen as
non--separability property) implies $\mathcal{F}>1/2$. So that entanglement
is a pre--requisite for using the state as a resource for the teleportation
of a coherent state.

There are two interesting regions in the plot that deserve some comments.
The first one, encompassing areas labelled as (\textit{III}) and (\textit{IV}%
) in the plot (\textit{lights green} and \textit{blue}), represents states
that violate the PHS bound ($w_{PHS}<0$) while they do not the Duan one ($%
w_{DUAN}\geq 0$). This apparent ambiguity can be solved noting that the 
\textbf{CM}s relative to these regions do not respect the condition (\ref%
{Duan_conditions}) so that the non--negativity of $w_{DUAN}$ does not imply
a separability of the state. On the other hand for such states $w_{PHS}<0$
implies that they are effectively entangled and that their density matrix $%
\rho $ cannot factorize into a convex combination of the tensor product of
density operators relative to the two different sub--systems. It is possible
to see that if these \textbf{CM}s are transformed by local squeezing
operations, as outlined in Ref. \cite{Duan00}, into a form that fulfills
conditions (\ref{Duan_conditions}), the transformed states show $w_{DUAN}<0$%
. We have numerically done a few tests on such \textit{odd} matrices,
verifying that once taken into that form, the states violate the Duan bound (%
\ref{Duan_necessary}). We note that, being the latter written in a more
general form, it is more useful from the practical point of view.

The second interesting region, labelled with (\textit{IV}) (\textit{light
green}) in Fig. \ref{Fig:criteria}, represents states that, although
entangled, cannot be used for teleporting coherent states. \textbf{CM}s
lying inside this area will not give $\mathcal{F}>1/2$. It is interesting to
note that such states fall also inside the region for which the Duan
criterion (\ref{DUANcr}) is not fulfilled. As above mentioned, once the
relative \textbf{CM}s are transformed into the form (\ref{CM_Duan}) by local
squeezing the transformed state will fulfill the Duan criterion in the form (%
\ref{Duan_necessary}) so that, in this new scenario, the system will be
entangled. At the same time, if this novel state is used as a quantum
resource for teleportation of a coherent state it will give $\mathcal{F}>1/2$
\cite{Pirandola06} so that local squeezings unveil entanglement. The initial
state lying in this area is entangled for PHS, being $w_{PHS}<0$, but, from
the point of view of teleportation, entanglement manifests itself in an
useless way. This entanglement can be made useful by locally transforming
the two subsystems.

We can see that the EPR criterion (region (\textit{I}), \textit{light brown}%
) offers a more restrictive condition with respect to the other two criteria
even for diagonal states.

Region (\textit{II}) (\textit{salmon}) represents the bound fixed by the
Duan criterion as a sufficient but not necessary condition. While region (%
\textit{V}) (\textit{white}) represent separable states.

\subsubsection{Mutual information and quantum discord\label{Sect:discord}}

Recent studies have shown that some particular classes of separable
correlated states, traditionally considered classical, show quantum features
useful for application in quantum technology \cite{Datta08}.

In statistical theory any correlation between two random variables $A$ and $%
B $ can be measured by their \textit{mutual information} defined by two
equivalent expressions:%
\begin{eqnarray}
I\left( A;B\right) &\equiv &H\left( A\right) +H\left( B\right) -H\left(
A,B\right)  \notag \\
&\equiv &H\left( A\right) -H\left( A|B\right) =H\left( B\right) -H\left(
B|A\right)  \label{mutual_classical}
\end{eqnarray}%
where $H\left( X\right) $ is the Shannon entropy (the classical analogue of
the $S\left( \rho \right) $, von Neumann entropy defined in Sect. \ref%
{Sect:purity}) associated to the statistical distribution of the variable $X$
($\left( X,Y\right) $ indicating the joint probability distribution and $%
\left( X|Y\right) $ the conditional distribution of $Y$ given the value of $%
X $). These two definitions, translated into the quantum language by
substituting $H$ with $S\left( \rho \right) $, the von Neumann entropy, do
not coincide anymore \cite{Giorda10}. So, while the translation of the first
one into the quantum language is straightforward and univocal, this is not
true for the second one.

The first of the two definitions (\ref{mutual_classical}), is unambiguously
referred to as the quantum mutual information \cite{Barnett89} between state 
$1$ and $2$ ($\rho $ representing the state of the bi--partite system as a
whole)%
\begin{equation}
\mathcal{I}\left( \rho \right) =S\left( \rho _{1}\right) +S\left( \rho
_{2}\right) -S\left( \rho \right) ~.  \label{quantum_mutual}
\end{equation}%
where $\rho _{1\left( 2\right) }=Tr_{2\left( 1\right) }\left[ \rho \right] $
are the partial traces. It relies only on the quantum state of the whole
system compared with the single sub--systems' states. In a bi--partite
system, described by a density matrix $\rho $, $\mathcal{I}(\rho )$
quantifies the total correlation between the subsystems $\rho _{1}$ and $%
\rho _{2}$. It is $>0$ for entangled states, while it is strictly $=0$ for
separable systems. It can be written in terms of the \textbf{CM} invariants
(see Sect. \ref{Sect:state}) and symplectic eigenvalues (see Eq. (\ref%
{symplectic}))%
\begin{equation}
\mathcal{I}\left( \mathbf{\sigma }\right) =f\left( n\right) +f\left(
m\right) -f\left( d_{+}\right) -f\left( d_{-}\right) ~,  \label{mutual}
\end{equation}%
with $f\left( x\right) $ given in Eq. (\ref{Funct_f}).

The second definition of Eq. (\ref{mutual_classical}), translated into the
quantum world, necessarily involves the conditional state of a subsystem
after a measurement performed on the other one. So that, the symmetry
between the two subsystems is broken. Since the conditional entropy $H\left(
A|B\right) $ requires to specify the state of $B$ given the state of $A$,
its definition, in quantum theory is ambiguous until the to--be--measured
observables on $A$ are selected so that the conditional state of $B$ can be
defined.

This discrepancy has lead to the concept of quantum discord $\mathcal{D}%
\left( \rho \right) $ \cite{Ollivier01}. A non zero $\mathcal{D}\left( \rho
\right) $ signals the presence of quantum features in the correlation
between the two sub--systems notwithstanding their separability or entangled
nature. So that $\mathcal{D}\left( \rho \right) $ is a measure of genuine
quantum correlation%
\begin{equation*}
\mathcal{D}\left( \rho \right) =\mathcal{I}\left( \rho \right) -\mathcal{C}%
\left( \rho \right) ,
\end{equation*}%
where $\mathcal{C}\left( \rho \right) $ is the amount of genuinely classical
correlation.

For a Gaussian state described by the \textbf{CM }(\ref{CM initial}), $%
\mathcal{D}$ becomes \cite{Giorda10}

\begin{equation}
\mathcal{D}\left( \mathbf{\sigma }\right) \mathcal{=}f(m)\mathcal{-}f\left(
d_{+}\right) -f\left( d_{-}\right) +f\left( \frac{n+2nm+2c_{1}c_{2}}{1+2m}%
\right) ~.  \label{discord}
\end{equation}%
We note that, as for the EPR criterion, in quantum discord there is an
asymmetry in the exchange of the two sub--systems. Again this is due to the
use of the concept of conditional states.

\section{Quantum markers evolution\label{Sect:evolved}}

As shown the \textbf{CM} of a bi--partite state undergoing to a lossy
transmission evolves as Eq. (\ref{CM post BS}). Consequently the different
quantum markers evolve as:%
\begin{eqnarray}
\mu _{T} &=&\frac{1}{4\sqrt{\text{det}\left[ \mathbf{\sigma }_{T}\right] }}~,
\notag \\
\mathcal{F}_{T} &=&\frac{1}{\sqrt{1+\left( m_{T}+n_{T}\right) ^{2}+2\left(
c_{2,T}-c_{1,T}\right) \left( 1+m_{T}+n_{T}\right) +2\left(
m_{T}+n_{T}-2c_{1,T}c_{2,T}\right) }}  \notag \\
w_{PHS,T} &=&n_{T}^{2}+m_{T}^{2}+2\left\vert c_{1,T}c_{2,T}\right\vert
-4\left( n_{T}m_{T}-c_{1,T}^{2}\right) \left( n_{T}m_{T}-c_{2,T}^{2}\right) 
\notag \\
w_{DUAN,T} &=&\sqrt{\left( 2n_{T}-1\right) \left( 2m_{T}-1\right) }-\left(
c_{1,T}-c_{2,T}\right)  \notag \\
w_{EPR,T} &=&n_{T}^{2}\left( 1-\frac{c_{1,T}^{2}}{n_{T}m_{T}}\right) \left(
1-\frac{c_{2,T}^{2}}{n_{T}m_{T}}\right) -\frac{1}{4}  \notag \\
\mathcal{I}_{T} &=&f\left( n_{T}\right) +f\left( m_{T}\right) -f\left(
d_{+,T}\right) -f\left( d_{-,T}\right)  \notag \\
\mathcal{D}_{T} &\mathcal{=}&f(m_{T})\mathcal{-}f\left( d_{+,T}\right)
-f\left( d_{-,T}\right) +f\left( \frac{n_{T}+2n_{T}m_{T}+2c_{1}c_{2}}{%
1+2m_{T}}\right)  \label{Ev_markers}
\end{eqnarray}%
where the subscript $_{T}$ indicates the quantity undergone to a lossy
transmission described by Eq.(\ref{CM post BS}).

We note that the vacuum state obtained for $T=0$ is a pure one \textit{i.e. }%
$\mu _{0}=1$. Moreover, in the ideal case (no loss), the OPO would generate
a pure state as well. Being $\mu _{T}<\mu _{0,1}$, for $T\neq 0,1$, the
evolution of $\mu _{T}$ is not monotonic. The purity of the composite system
cannot be considered a general entanglement marker \cite{Adesso04}. As a
matter of fact, any pair of physical systems in a pure state have $\mu =1$
even if they are disentangled. On the other hand, in our specific case,
having a precise hypothesis on the ideal state (a pure twin--beam diagonal
state) outing the OPO crystal allows us to consider $\mu $ as a measure of
the total decoherence that has affected the state.

It is easy to see that $\mathcal{F}_{T}$, $w_{PHS,T}$, and $w_{DUAN,T}$
describe properties very robust under decoherence. Once $\mathcal{F}>1/2$, $%
w_{PHS}<0$ and $w_{DUAN}<0$ for $T=1$ they will keep breaking the respective
bounds for every level of loss. Both mutual information and quantum discord
show, with respect to loss, the same feature even if decoherence affects
their amount. On the contrary a state that show EPR--like correlation ($%
w_{EPR}<0$) for $T=1$ will not keep this property along the propagation so
that some particular protocol based on this property cannot be reproduced.
Under a total loss greater than $50\%$ any state looses this correlation
property.

\section{The Experiment\label{Sect:Experiment}}

The transmission over a Gaussian channel is described by Eq. (\ref%
{Fokker-Planck}). As already mentioned this evolution is in all equivalent
to a fixed amount of loss introduced by a fictitious beam--splitter. The
actual experimental apparatus is made of the CV entangled state source, a
variable attenuator (mimicking the BS), and a state characterization stage.
A block diagram of the experimental setup is presented in Fig. \ref%
{Fig:setup}

\begin{figure}[tph]
\includegraphics[width=0.48\textwidth]{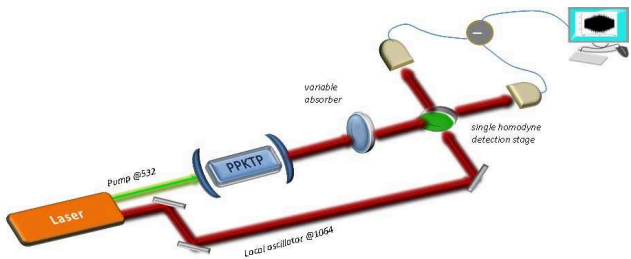}
\caption{\textit{(color online)} Block diagram of the experimental setup.
The details on the OPO source are given in Ref. \protect\cite{APB08}, while
the characterization stage, based on a single homodyne detector, is fully
described in Ref. \protect\cite{JOB05}.}
\label{Fig:setup}
\end{figure}

\subsection{CV entangled state source}

The state source relies on a CW internally frequency doubled Nd:YAG laser
(Innolight Diabolo) whose outputs @532nm and @1064nm are respectively used
as the pump for a non degenerate OPO and the local oscillator (LO) for the
homodyne detector. The OPO is set to work below the oscillation threshold
and it provides at its output two entangled thermal states (the signal $a$
and the idler $b$).

The OPO is based on an $\alpha $-cut periodically poled KTP non linear
crystal (PPKTP, \textit{Raicol Crystals Ltd}. on custom design) which allows
implementing a type II phase matching with frequency degenerate and cross
polarized signal and idler beams, for a crystal temperature of $\approx $ 53$%
^{\circ }$C. The transmissivity of the cavity output mirror, $T_{out}$, is
chosen in order to guarantee, together with crystal losses ($\kappa $) and
other losses mechanisms ($T_{in}$), an output coupling parameter $\eta
_{out}=T_{out}/(T_{in}+\kappa )$ @1064 nm of $\approx $ $0.73$,
corresponding to an experimental linewidth of $15$ MHz @1064 nm. In order to
obtain a low oscillation threshold, OPO cavity geometry is set to warrant
simultaneous resonance on the pump, the signal and the idler \cite{APB08}.
Measured oscillation threshold is around $50$ mW; during measurement runs
the system has been operated at $\approx $60\% of the threshold power to
avoid unwanted non--Gaussian effects \cite{OPEX05}.

The two beams outing the OPO are transmitted through a filter of variable
optical density that mimics the BS. The loss level introduced by the filter
is polarization independent and can be tuned from a few percent up to more
than 99\%.

\subsection{Characterization stage}

The signal and idler modes are then sent to the covariance matrix
measurement set-up: this consists in a preliminary polarization system, that
allows choosing the beam to be detected and a standard homodyne detector.
The polarization system is made of an half-wave plate ($\lambda /2$)
followed by a polarizing beam splitter (PBS); the different wave-plate
orientations allow choosing the beam to be transmitted by the PBS: the
signal ($a$), the idler ($b$) or their combinations $c=\frac{1}{\sqrt{2}}%
(a+b)$ or $d=\frac{1}{\sqrt{2}}(a-b)$. Two other combinations, $e=\frac{1}{%
\sqrt{2}}(ia+b)$ and $f=\frac{1}{\sqrt{2}}(ia-b)$, can be obtained by
inserting before the PBS an additional quarter wave plate ($\lambda /4$) 
\cite{JOB05}. Acquisition times are considerably short thank to pc-driven
mechanical actuators that allow setting the $\lambda /2$ and $\lambda /4$
positions in a fast and well calibrated manner.

Once a beam is selected, it goes to an homodyne detector put downstream the
PBS. This exploits, as local oscillator, the laser output @1064 nm,
previously filtered and adjusted to match the geometrical properties of the
OPO output: a typical interferometer visibility is $0.98$. The LO phase $%
\theta $ is spanned to obtain a $2\pi $ variation in $200$ ms. In order to
avoid low frequency noise the homodyne current is demodulated at $\Omega $=3
MHz and low-pass filtered ($B$=$300$ KHz). Then it is sampled by a PCI
acquisition board obtaining $10^{6}$pts/run, with 14-bit resolution. The
electronic noise floor is 16 dBm below the shot noise level, corresponding
to SNR $\approx $40. Data are analysed by a \copyright Mathematica routine
that extract from the six homodyne traces the relevant quadrature variances
necessary for reconstructing the whole covariance matrices \cite{JOB05}.

\section{Experimental results\label{Sect:Results}}

We have performed different sets of measurement under lossy transmission in
order to investigate different loss regimes. Each experimental \textbf{CM}
comes from seven homodyne traces: the shot noise calibration trace, obtained
by obscuring the OPO output, six traces each for one of the six modes
required for a full state characterization. Contrarily to other previous
experiments, where quantum tomographic routine were used in order to
retrieve experimental \textbf{CM}s \cite{PRL09}, we have evaluated \textbf{CM%
}s by a simpler \copyright Mathematica routine that calculates relevant
second order moments of homodyne distributions in a faster way without
enhancing the experimental indeterminacy on the \textbf{CM} elements. We
have tested on a few \textbf{CM}s that this procedure gives results in all
compatible with quantum tomography. We have also checked, with the standard
procedure outlined in \cite{PRA09}, that the states under scrutiny were
effectively Gaussian.

Once a \textbf{CM} is obtained the different entanglement witnesses ($%
w_{PHS} $, $w_{DUAN}$, and $w_{EPR}$ Eqs. (\ref{witnesses})), state purity ($%
\mu $ Eq. (\ref{purity})), teleportation fidelity ($\mathcal{F}$ Eq. (\ref%
{Fidelity_th})), quantum discord ($\mathcal{D}$ Eq. (\ref{discord})), and
mutual information ($\mathcal{I}$ Eq. (\ref{quantum_mutual})) are
calculated. Then, the overall decoherence, \textit{i.e. }the total level of
loss that includes OPO cavity escape efficiency, propagation loss, filter
absorption, homodyne efficiency, is assigned as a label to the measurement 
\cite{Bowen03}. This expected level of decoherence is then compared to the
theoretical one obtained by inverting Eq. (\ref{CM post BS}) and solving for 
$T$ under the condition $\det (\mathbf{\sigma }_{1})=1/16$; thus requiring
that $\mathbf{\sigma }_{1}$ represents a pure state (see the discussion at
the end of Sect. \ref{Sect:state}). We have verified that, even if
experimental \textbf{CM}s do not reproduce exactly \textit{diagonal }states,
all of them respect the Duan conditions (\ref{Duan_conditions}) within
experimental indeterminacies. So that for the analysed matrices the Duan
witness $w_{DUAN}$ represents a sufficient and necessary condition for
entanglement.

In all the reported plots we have considered the less decohered datum
(obtained for $T=0.63$) as a reference so that all the reported theoretical
curves are obtained imposing that Eqs. (\ref{CM_elements_evolution}) and (%
\ref{Ev_markers}) evaluted for $T=0.63$ give the measured values.

The total losses we have measured span the interval $37-99\%$ ($0.01\leq
T\leq 0.63$). We note that $T=0.63$, in absence of extra loss and having a
cavity escape efficiency of $\approx 0.73$, implies an overall state
detection efficiency of $\approx 0.86$ in agreement with an homodyne
visibility of $0.98\pm 0.01$, a photodiode (nominal) efficiency of $0.90\pm
0.01$ and residual transmission loss between the OPO output mirror and the
detector surface of $0.01\pm 0.01$.

\begin{figure}[!tph]
\includegraphics[width=0.48\textwidth]{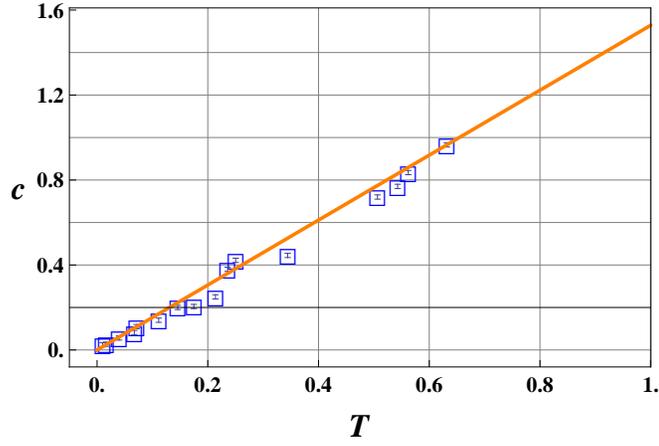}
\caption{\textit{(color online)} behaviour of the averaged correlation term $%
\left\vert c_{1,T}\right\vert +\left\vert c_{2,T}\right\vert /2$ in Eq. (%
\protect\ref{CM post BS}). As expected the correlation reduces linearly with 
$T$. The full (\textit{dark orange}) line represents the linear behaviour
calculated starting from the first experimental point we have measured ($%
T=0.63$). Error bars are smaller than data points and amount to $\pm 0.01$.}
\label{Fig:c}
\end{figure}

In Fig. \ref{Fig:c} we report the behaviour vs. $T$ of the averaged
correlation term ($\left( \left\vert c_{1,T}\right\vert +\left\vert
c_{2,T}\right\vert \right) /2$ see Eq. (\ref{CM_elements_evolution})). As
expected the correlation between the two sub--systems degrades linearly with
the total loss ($T\rightarrow 0$). The expected behaviour (full \textit{dark
orange} line), obtained by considering the less absorbed \textbf{CM} ($%
T=0.63 $) as a reference, follows quite well the reported data. Actually,
data refer to acquisition taken on different days so that, the scattering of
the point around that line is more due to source long--term dynamics then to
actual deviation from the Lindblad model. At the same time the fact that the
points are reasonably close to that line proves that the long term stability
of the source can be considered quite good.

\begin{figure}[tph]
\includegraphics[width=0.48\textwidth]{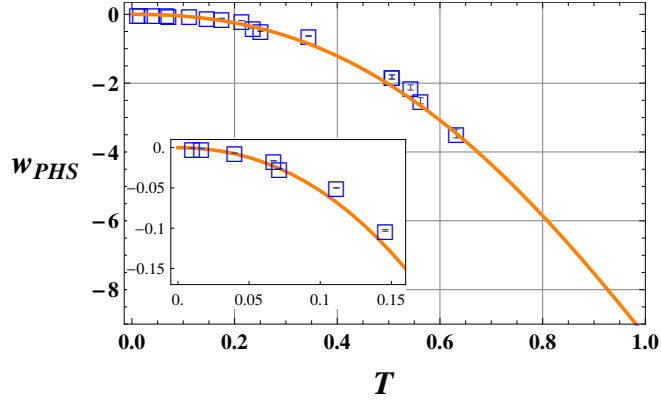}
\caption{\textit{(color online)} $w_{PHS}$ vs. $T$. The full (\textit{dark
orange}) line represents the expected behaviour calculated by the third of
Eqs. (\protect\ref{Ev_markers}) setting the first experimental point at $%
T=0.63$ as the intial datum. Error bars, obtained by propagating the
experimental indeterminacies in Eq. (\protect\ref{witnesses}a), range
between $10^{-4}$ and $0.1$. In the inset we report the high loss regime ($%
T<0.15$) for better enlightning the un--separability, as witnessed by $%
w_{PHS}$, even in presence of strong decoherence.}
\label{Fig:PHS}
\end{figure}

\begin{figure}[tph]
\includegraphics[width=0.48\textwidth]{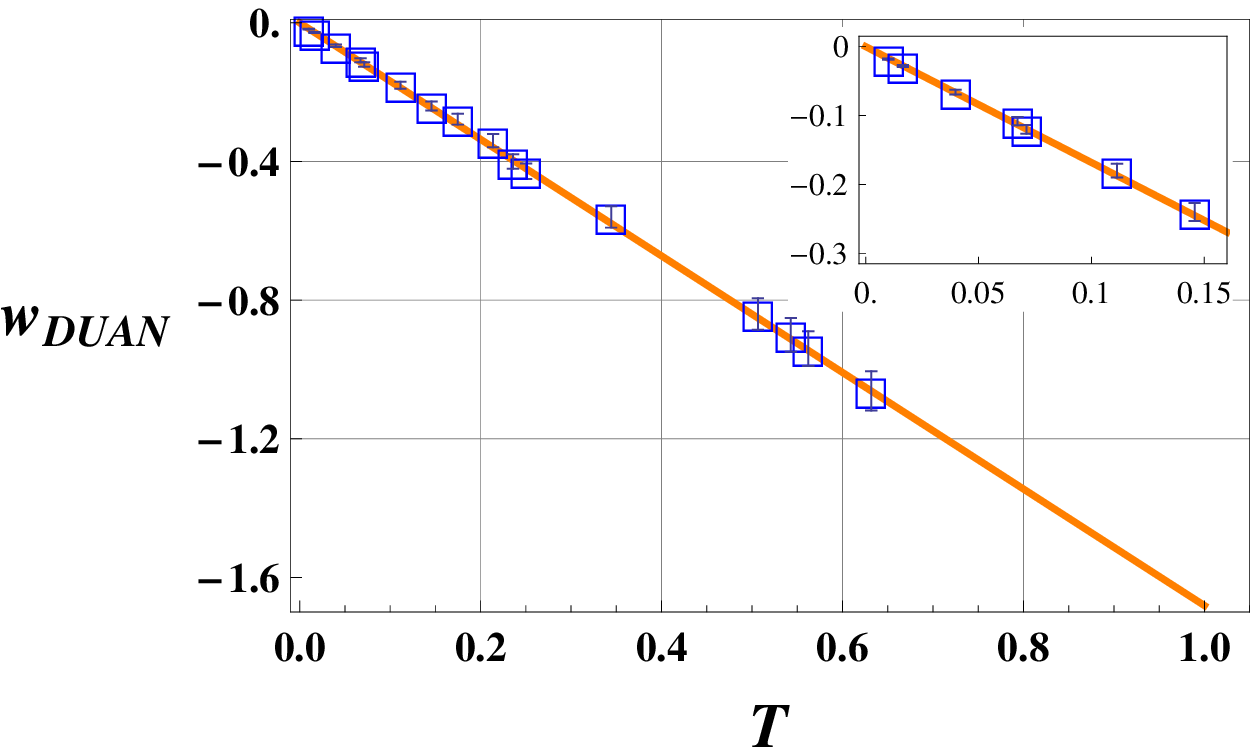}
\caption{\textit{(color online)} $w_{DUAN}$ vs. $T$. The full (\textit{dark
orange}) line represents the expected behaviour calculated by the fourth of
Eqs. (\protect\ref{Ev_markers}) setting the first experimental point at $%
T=0.63$ as the intial datum. Error bars, obtained by propagating the
experimental indeterminacies in Eq. (\protect\ref{witnesses}b), range
between $0.01$ and $0.06$. In the inset we report the plot for $T<0.15$ in
order to better visualize the persistence of entanglement, as witnessed by $%
w_{DUAN}$, even in presence of strong decoherence.}
\label{Fig:DUAN}
\end{figure}

As already mentioned, $w_{PHS}$ and $w_{DUAN}$ describe a physical property
of the state that is strong under decoherence as proved for lower loss
(below 90\%) in Ref. \cite{Bowen03}. They are symptoms of un--separability,
in the sense that the system state cannot be described by a density matrix
in the form of Eq. (\ref{SepSt}) \cite{Werner89}. This can be seen in Figs. %
\ref{Fig:PHS} and \ref{Fig:DUAN} where $w_{PHS}$ and $w_{DUAN}$ are plotted
vs. $T$. We have also enlarged, in the insets the region for strong loss ($%
T<0.15$) to prove that, even if the analysed state is very close to a two
mode vacuum (the total average number of photon ($\left( n+m-1\right) /2$)
reduces to $0.02\pm 0.01$) it is still experimentally possible to prove that
the state is non--separable. It has to be noted that, while, for $%
T\rightarrow 0$, $w_{PHS}$ approaches its classical limit non--linearly (see
the third of Eqs. (\ref{Ev_markers})), $w_{DUAN}$ (see the fourth of Eqs. (%
\ref{Ev_markers})) is linear. Thus, in the very high loss regime it becomes
more reliable to assess entanglement using the latter than the former.

\begin{figure}[tph]
\includegraphics[width=0.48\textwidth]{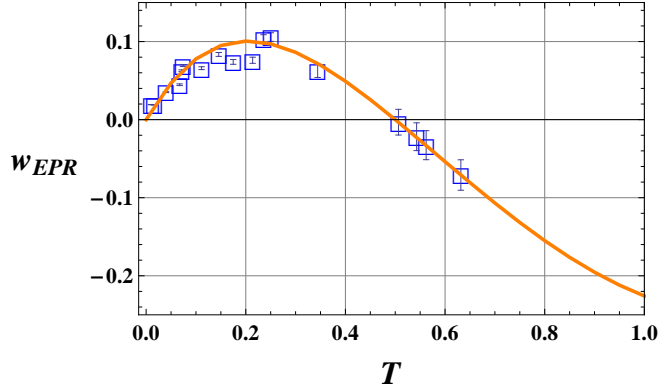}
\caption{\textit{(color online)} $w_{EPR}$ vs. $T$. The full (\textit{dark
orange}) line represents the expected behaviour calculated by the fifth of
Eqs. (\protect\ref{Ev_markers}) setting the first experimental point at $%
T=0.63$ as the intial datum. Error bars, obtained by propagating the
experimental indeterminacies in Eq. (\protect\ref{witnesses}c), range
between $2\times 10^{-4}$ and $0.02$. They are considerably larger for point
at low losses. As expected for total losses larger than $0.5$ $w_{EPR}>0$
and the state does not show \textit{EPR} correlation.}
\label{Fig:EPR}
\end{figure}

$w_{EPR}<0$ indicates that the state exhibits EPR--like correlation so that
it is possible to gain information on the expectation value of one
observable on one sub--system with a precision better than the standard
quantum limit once the EPR companion is measured on the other sub--system.
This feature is by far the most fragile under decoherence: for $T<0.5$ no
state can keep this quantum feature. This can be understood from the fact
that loss, a stochastic process, affects directly the degree of correlation
between the two sub--systems while the system representation (\textit{i.e.}
its un--separability) is only smoothed by this process. It is relevant to
note that $T=0.5$ also corresponds to the minimum state purity $\mu $. So
that, loosing the EPR character coincides with the maximum mixedness for the
state during its propagation. In Fig. \ref{Fig:EPR} we report the
experimental behaviour found for $w_{EPR}$ for our system. Measured \textbf{%
CM}s for $T<0.5$ all show $w_{EPR}>0$. A positive $w_{EPR}$ indicates that,
for these states, any attempt to gain information on one sub--system by
measuring the other would result less precise than the standard quantum
limit.

\begin{figure}[tph]
\includegraphics[width=0.48\textwidth]{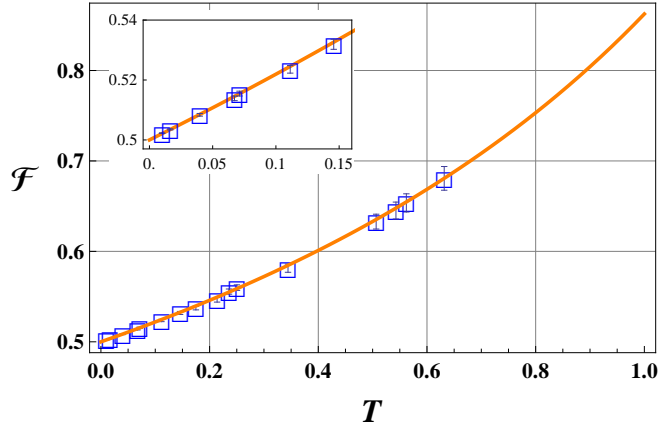}
\caption{\textit{(color online)} $\mathcal{F}$ vs. $T$. The full (\textit{%
dark orange}) line represents the expected behaviour calculated by the
second of Eqs. (\protect\ref{Ev_markers}) setting the first experimental
point at $T=0.63$ as the intial datum. Error bars, obtained by propagating
the experimental indeterminacies in Eq. (\protect\ref{Fidelity_th}), range
between $10^{-4}$ and $0.01$. In the inset we report the plot for $T<0.15$
in order to underline the persistence of a quantum teleportation regime even
in presence of strong decoherence (high loss) thus proving that even for, in
principle, infinite distance this class of states would allow to perform the
teleportation of a coherent state with a fidelity above $1/2$.}
\label{Fig:fid}
\end{figure}

An important signature for an entangled CV state is its ability of acting as
a quantum resource in the CV teleportation protocol for coherent state. In
Eq. (\ref{Fidelity_th}) we have expressed the fidelity $\mathcal{F}$ as a
function of the \textbf{CM} elements. $\mathcal{F}$, as $w_{PHS}$ and $%
w_{DUAN}$, represents a robust signature of quantum properties for the state
undergoing to a lossy transmission. In particular, in Fig. \ref{Fig:fid}, we
see that even in the high loss regime, $\mathcal{F}$ remains above the
classical limit of $0.5$ (see the inset for greater details). Thus proving
that CV entangled state, as the one produced by our source, could be used as
resource for realising teleportation protocol of coherent state, in
principle, at an infinite distance.

Eventually we have retrieved, from our \textbf{CM}s, the value for the
quantum mutual information $\mathcal{I}\left( \mathbf{\sigma }\right) $ (Eq.
(\ref{mutual})) and quantum discord $\mathcal{D}\left( \mathbf{\sigma }%
\right) $ (Eq. (\ref{discord})).

\begin{figure}[tph]
\includegraphics[width=0.48\textwidth]{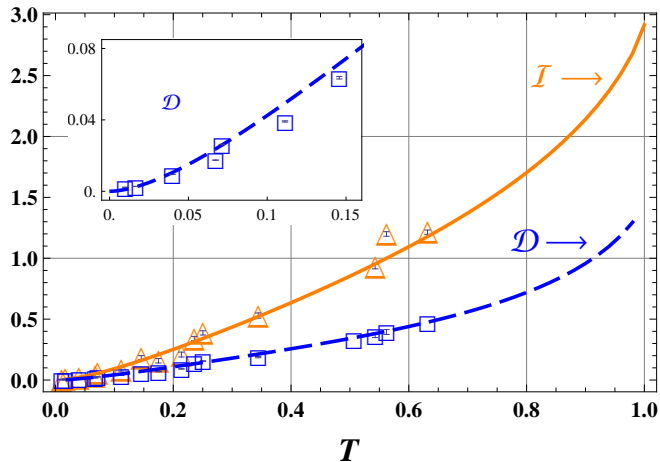}
\caption{\textit{(color online)} $\mathcal{I}$ and $\mathcal{D}$ vs.$T$. The
full (\textit{dark orange}) and dashed (\textit{blue}) lines represent the
expected behaviours calculated by the sixth and seventh of Eqs. (\protect\ref%
{Ev_markers}) setting the first experimental point at $T=0.63$ as the intial
datum. Error bars, obtained by propagating the experimental indeterminacies
in Eqs. (\protect\ref{mutual}) and (\protect\ref{discord}), respectively,
range between $3\times 10^{-3}$ and $0.02$ for $\mathcal{I}$ and $10^{-4}$
and $0.03$ for $\mathcal{D}$. In the inset we report the $\mathcal{D}$ data
for $T<0.15$ in order to underline the persistence of true quantum
correlation even in presence of strong decoherence (high loss). Note that
the data for $\mathcal{I}$ scatter more from the expected behaviour may be
signalling extra classical correlations.}
\label{Fig:discord}
\end{figure}

In Fig. \ref{Fig:discord} we report the experimental data together with the
expected behaviours, as usually calculated considering the less decohered
matrix as a reference, for $\mathcal{I}$ and $\mathcal{D}$ vs. $T$. As it
can be seen the quantum discord follows very well its \textit{theoretical}
line while quantum mutual information is a little more scattered around it.
Moreover, our data prove that even in presence of strong decoherence, it is
possible to evaluate that $\mathcal{D}$ keeps $>0$, within the experimental
indeterminacies, all the way down to an highly absorbed state. We note that\
Gaussian quantum discord is attracting, very recently, a lot of experimental
interest \cite{Gu12,Blandino12,Madsen12}. In particular, in Ref. \cite{Gu12}%
, the authors give an operational significance to quantum discord as the
possibility of encoding quantum information in separable states. In Ref. 
\cite{Blandino12} the optimal strategy for evaluating $\mathcal{D}\left( 
\mathbf{\sigma }\right) $ in homodyne measurement is presented. It is
interesting to compare our experimental plot with the one reported in Ref. 
\cite{Madsen12} where the authors analyse the quantum discord under the
lossy transmission of one of the two sub--systems. We note that in their
case the scattering of the experimental points around the theoretical curve
is almost equivalent for $\mathcal{I}$ and $\mathcal{D}$ while in our case
there is a clear difference.

\section{Conclusions\label{Sect:conclusions}}

Gaussian bi--partite states are one of the most renown resources for
implementing CV quantum communication protocols such as CV teleportation of
coherent states. In this paper we have experimentally analysed how
decoherence affects different entanglement criteria and quantum markers for
a CV bi--partite state outing a sub--threshold type--II OPO. The decoherence
is experimentally introduced by transmistting the quantum state through a
variable attenuator. Before illustrating our experimental results we have
discussed in details the relationship between the three different
entanglement criteria used in the CV\ framework and linked them to the
teleportation fidelity and quantum discord. The latters represent two
possible quantum signatures for evaluating the ability of this class of
states in quantum communication protocols.

On one hand, our findings prove that the Lindblad approach for describing
lossy transmission is valid all the way down to strongly decohered states.
On the other hand, with this paper, we prove that the particular class of
states we have analysed keeps, within the experimental indeterminacies, its
main quantum signatures, \textit{i.e. }the possibility of realizing quantum
teleportation of coherent states with a fidelity above $0.5$ and a quantum
discord above $0$ for a total loss of $\approx 99\%$. This proves that the
class of CV entangled states, we analysed, would allow, in principle, to
realize quantum teleportation over an infinitely long Gaussian channel.

In analysing how quantum discord (see Fig. \ref{Fig:discord}) and quantum
mutual information behave under decoherence we interestingly found that the
scattering of the points around the theoretical curve is significantly more
evident for the quantum mutual information may be signalling that a key
role, in our case, is played by unexpected classical correlations. This
point will be subject of further theoretical and experimental
investigation.\medskip

The authors thank S. De Siena and F. Illuminati for useful suggestions and
discussions on the theoretical aspects.

\end{document}